\documentclass[sigconf]{acmart}

\usepackage{xspace}
\usepackage{subcaption}
\usepackage{caption}
\usepackage{graphicx}
\usepackage{multirow}
\usepackage{rotating}
\usepackage{makecell}
\usepackage{booktabs}
\usepackage{diagbox}
\usepackage{xcolor}
\usepackage{enumitem}

\usepackage[show]{chato-notes}

\newcommand{\algoname}[1]{{\sc #1}\xspace}
\newcommand{\dataset}[1]{{\sf #1}\xspace}
\newcommand{\msna}{\dataset{MSN-1}}
\newcommand{\msnab}{\dataset{\textbf{MSN-1}}}
\newcommand{\istella}{\dataset{Istella}}
\newcommand{\istellab}{\dataset{\textbf{Istella}}}

\newcommand{\QS}{\algoname{QuickScorer}}
\newcommand{\mart}{\algoname{MART}}
\newcommand{\lmart}{\algoname{$\lambda$-Mart}}

\newcommand{\x}{$\times$}
\newcommand{\dm}{$\Delta$}
\newcommand{\ra}[1]{\renewcommand{\arraystretch}{#1}}

\newcommand{\lear}{\algoname{LEAR}}
\newcommand{\learb}{\algoname{\textbf{LEAR}}}
\newcommand{\ept}{\algoname{EPT}}
\newcommand{\eptb}{\algoname{\textbf{EPT}}}

\newcommand{\rp}[1]{\textcolor{magenta}{#1}}

\copyrightyear{2021}
\acmYear{2021}
\setcopyright{none}
\acmConference[SIGIR '21]{Proceedings of the 44th International ACM SIGIR Conference on Research and Development in Information Retrieval}{July 11--15, 2021}{Virtual Event, Canada}
\acmBooktitle{Proceedings of the 44th International ACM SIGIR Conference on Research and Development in Information Retrieval (SIGIR '21), July 11--15, 2021, Virtual Event, Canada}
\acmDOI{}
\acmISBN{}
\begin{document}
\fancyhead{}

\title{Learning Early Exit Strategies for Additive Ranking Ensembles}

\author{Francesco Busolin}
\affiliation{%
  \institution{Ca' Foscari University of Venice, Italy}
}
 \email{francesco.busolin@unive.it}

\author{Claudio Lucchese}
\affiliation{%
  \institution{Ca' Foscari University of Venice, Italy}
}
 \email{claudio.lucchese@unive.it}

\author{Franco Maria Nardini}
\affiliation{%
  \institution{ISTI-CNR, Pisa, Italy}
}
 \email{francomaria.nardini@isti.cnr.it}

\author{Salvatore Orlando}
\affiliation{%
  \institution{Ca' Foscari University of Venice, Italy}
}
 \email{orlando@unive.it}

\author{Raffaele Perego}
\affiliation{%
  \institution{ISTI-CNR, Pisa, Italy}
}
 \email{raffaele.perego@isti.cnr.it}

\author{Salvatore Trani}
\affiliation{%
  \institution{ISTI-CNR, Pisa, Italy}
}
 \email{salvatore.trani@isti.cnr.it}

% !TEX root = paper.tex
% !TeX spellcheck = en_US
\begin{abstract}
Modern search engine ranking pipelines are commonly based on large machine-learned ensembles of regression trees. 
%The tight constraints on query response time require to devise solutions for making document scoring as fast as possible.
%In this paper, 
We propose \lear, a novel -- learned -- technique aimed to reduce the average number of trees traversed by documents to accumulate the scores, thus reducing the overall query response time. \lear exploits a classifier that predicts whether a document can early exit the ensemble because it is unlikely to be ranked among the final top-$k$ results. The early exit decision occurs at a sentinel point,
%in the forest
i.e., after having evaluated a limited number of trees, and the partial scores are exploited to filter out non-promising documents.
We evaluate \lear by deploying it in a production-like setting,  adopting a state-of-the-art algorithm for ensembles traversal. 
We provide a comprehensive experimental evaluation on two public datasets. The experiments show that \lear has a significant impact on the efficiency of the query processing without hindering its ranking quality. In detail, on a first dataset, \lear is able to achieve a speedup of 3$\times$ without any loss in NDCG@10, while on a second dataset the speedup is larger than 5$\times$ with a negligible NDCG@10 loss ($<0.05$\%).
\end{abstract}


\maketitle

% !TEX root = paper.tex
% !TeX spellcheck = en_US

\section{Introduction}
\label{sec:intro}
Query processors of modern search engines rely on sophisticated ranking pipelines aimed at optimizing precision-oriented metrics at small cutoffs.  Learning-to-rank (LtR) techniques are commonly used to train complex models able to precisely re-rank a set of candidate documents. 
% retrieved from the index for each query and compute the set of  top-$K$ results returned to the user.
%Among the s
State-of-the-art solutions include  
%for candidate re-ranking, particular importance have the models based on 
additive ensembles of regression trees, such as \mart~\cite{Friedman00greedyfunction} and \lmart~\cite{lambdamart,burges2010ranknet}, learned by gradient boosting algorithms.
%Unfortunately the most effective LtR solutions are ensembles of hundreds of regression trees, such as
%Lambda-MART (\lmart)~\cite{lambdamart,burges2010ranknet} and Multiple Additive Regression Trees (. 
Since such ensemble are made of hundreds of additive regression trees,  the tight constraints on query response time require suitable solutions able to provide an optimal  trade-off between document scoring time and ranking effectiveness~\cite{Capannini20161161}.
%, given that the accuracy of  ranking ensembles is highly correlated with the number of trees in the model.
%quickly score documents thus reducing the processing time of such huge LtR tree ensembles, without hindering their ranking effectiveness.
%In the literature, several techniques have been proposed to address this issue.  
Among the main contributions in the area, we cite the algorithms for the efficient traversal of tree ensembles \cite{rapidscorer18,Dato2016,quickscorer15}. Alternative methods are concerned with strategies for pruning the ensemble during or after the training phase~\cite{lucchese2017x,xcleaver,Lucchese2016}, and budget-aware learning-to-rank algorithms \cite{Wang2010,asadi2013training}. 
Furthermore, researchers investigated early termination heuristics aimed to reduce, on a document- or query-level basis, the cost of the scoring process \cite{EarlyBarla10, QLEE2020, berxit}. These works studied  the impact of the proposed early termination strategies on both the latency and ranking accuracy.
Finally, other techniques looked at a composite scenario, where feature extraction costs and system effectiveness are balanced across multiple re-ranking stages \cite{RoiSIGIR17}. 

In this paper, we investigate document-level \textit{early exit} strategies for additive ranking ensembles by generalizing and building upon the state-of-the-art method introduced by Cambazoglu \emph{et al.}~\cite{EarlyBarla10}, who  
%It is worth mentioning that early exit is a (sometimes unsafe) strategy very common in IR, in particular to make more efficient the initial stages of query processing pipeline, by limiting the number of postings processed in inverted file indexes \rp{\cite{efficientWSE}}.
%In additive ranking ensembles, where documents are re-ranked by traversing the model to accumulate their final scores, Cambazoglu \emph{et al.} 
proposed some heuristic techniques to force documents to early exit the ensemble if they are unlikely to be included in the top-$k$ results. %, thus %. The introduction of document early exit allows to 
%speeding up  query processing. 
%In this paper, 
%Unlike the heuristic methods proposed in~\cite{EarlyBarla10},
%We devise a machine learning (ML) solution for early terminating document scoring. Specifically, 
We devise a machine learning framework (ML), called \lear (\textit{Learned EArly exit Ranking}), for early terminating document scoring. \lear is based on a binary classifier that exploits query-document features along with their score/rank cumulated up to a given ensemble's tree, called ``sentinel''. The classifier predicts whether the document should exit the ensemble because it will unlikely be ranked among the top-$k$ ones or it should continue the traversal of the rest of the ensemble.
%\fquestion{oppure perche' trattasi di documento irrilevante che se finisse nei top-k ridurrebbe la qualita' del ranking} 
We provide an in-depth study of the possible solutions to training effective early-exiting classifiers, we discuss their accuracy and their placements in the ensemble at specific sentinels\footnote{Code available at \url{https://github.com/hpclab/learning-exit-strategies-ensembles}}. We also provide an analysis of the speedup achieved by the introduction of \lear in \QS, a state-of-the art algorithm for ensemble traversal~\cite{quickscorer15}.
%\rp{Finally,} we evaluate our technique by deploying it in a production-like \rp{scenario} using \QS, a state-of-the-art algorithm for \rp{ensemble traversal}~\cite{quickscorer15}. 
%This algorithm can 
%The scoring algorithm exploits novel data structures to store the tree ensemble and a novel method that traverse the ensemble feature by feature. Indeed, the algorithm 
%manage  ensembles partitioned in blocks of trees to optimize the use of \rp{the fastest levels of the} memory hierarchy.
%\rp{We assess} that such block boundaries are a suitable point where a sentinel can be placed to activate our early exit classifier. 
%In conclusion, we combine a fast algorithm to traverse tree ensembles along with an ML-based technique aimed to reduce the number of trees each document has to traverse to be scored. 
%We provide a comprehensive analysis of the performance of our learned early exit strategy integrated in the \QS algorithm.
The experiments, conducted on two well-known public LtR datasets, namely \msna and \istella, %, i.e., \msna and \istella. The results 
show that our learned solution for document-level early exit provides up to 5$\times$ speedup with a negligible loss, lower than 0.05\%,  in terms of NDCG@10 and significantly outperforms the previous state-of-the-art solution by Cambazoglu \emph{et al.}~\cite{EarlyBarla10}.

% !TEX root = paper.tex
% !TeX spellcheck = en_US

\section{Learning early exit strategies}
\label{sec:learning}

%\ftodo{qui discutiamo l'idea: classificare documenti determinando la probabilità di essere nella lista top-k finale. problema di binary classification dove si usa la probabilità di essere 1. variabilità del threshold sulla probabilità.}

%\section{Document Early-Exit Framework}
In this section, we discuss \lear, our method to force documents to early exit (EE) the ensemble. 
To motivate and introduce our contribution,
%this further step in the EE framework applied to additive ensembles of decision trees, 
we first evaluate the main limits of the heuristics proposed by Cambazoglu \emph{et al.}~\cite{EarlyBarla10}.

\medskip
\noindent\textbf{Heuristic EE techniques.} 
%Cambazoglu \emph{et al.} in their paper propose 
Four EE heuristic techniques are proposed in \cite{EarlyBarla10}, where the best are \textit{EE Using Rank Thresholds} (ERT) and \textit{EE Using Proximity Thresholds} (EPT).
At a given sentinel $s$ in the ensemble, ERT sorts the candidate documents for a given query in decreasing order of their partial scores computed by evaluating the first $s$ trees. Then, given a pre-tuned rank threshold $k_s$, only the top-$k_s$ documents are evaluated by the remainder of the ensemble, while the other documents maintain their ranking at the sentinel without any additional computation.
%ERT sorts the query's documents in decreasing order of their scores, thus determining the document ranks. The documents having a rank larger than an offline-computed rank threshold $k$ early exit the ensemble, and are no longer scored by the rest of trees.
Besides ranks, EPT  exploits document ranks to select also those document in proximity of the top-$k_s$. Let $\sigma_{k_s}$ be the partial score of the $k_s$-th best document at the sentinel, all documents with a score smaller than $\sigma_{k_s} - p$ early exit the ensemble, where $p$ is a fine-tuned proximity threshold.

%Besides ranks, EPT also exploits the scores of the query's documents for EE decisions. For each query, at sentinel $s$ the score of the document at rank $k$ is collected.  Then,not only the first $k$ documents are kept, but also documents that are within a score proximity of the $k$-th. 
%Therefore, EPT is weaker than ERT in terms of the number of document that are forced to early exit the ensemble. Since less documents stop scoring at the sentinel $s$, the efficiency of EPT is smaller that EPT (i.e., speedups against the full method without EE decreases), but the effectiveness hopefully increases (i.e., NDCG@10 increases, and approaches the quality measure obtained by the full  method without EE).
%
%CLA: vero solo a parità di k ma non mi pare scontato
%
The rationale of EPT is to avoid a short-sighted top-$k_s$ selection when several documents have close scores and therefore they are equally likely to be ranked among the best results by the whole ensemble. In the case of such uncertain queries, EPT allows a larger number of candidate documents to be selected. 

To motivate \lear, i.e., our novel ML-based EE technique, we evaluate ERT and EPT when applied to a \lmart additive ensemble made up of $1{,}047$ trees and trained on the \msna
%MSLR-WEB30K
\footnote{\url{https://www.microsoft.com/en-us/research/project/mslr/}} dataset (fold 1) using the LightGBM\footnote{\url{https://github.com/Microsoft/LightGBM}} library.   Table \ref{tab:motivation} compares the  performance figures -- in terms of NDCG@10 and speedup --   obtained by the following methods on the test set: 
\begin{itemize}[leftmargin=*]
\item a ranking that exploits the complete ensemble without EE (\textit{Full});
\item the ideal EE strategy (EE$_{ideal}$), where for each query $q$ an oracle predicts the optimal value for $k_s^q$, where all the documents of rank greater than $k_s^q$ at the sentinel $s$ can safely early exit the ensemble. The value $k_s^q$ is the minimum one to guarantee that the measure NDCG@10 for query $q$ is the same as the original one (\textit{Full}), i.e., we select a distinct cut $k_s^q$ per each query that guarantees that all the top-10 documents, appearing at the end of the original whole ensemble, are kept in the ensemble and continue to score. 
\item two versions of ERT, with $k_s=15$ and $k_s=20$. We experimentally fine-tuned $k_s$ and these values resulted the best performing ones;
\item two versions of EPT, where we use $k_s=15$ and two values of score proximity $p$ ($p=0.2$ and $p=0.5$). Larger values of $p$ imply less documents that are stopped at the sentinel.
\end{itemize}

\begin{table}[tb]
%\small
    \def\arraystretch{0.9}
    \setlength{\tabcolsep}{0.9mm}
    \caption{Best strategies in~\cite{EarlyBarla10} vs Full and EE$_{ideal}$. 
    %Comparison of the traditional method without EE (\textit{Full}) with ERT, EPT, and EE$_{ideal}$ at sentinel $s=50$.
    \label{tab:motivation}}
    \centering
    \ra{1.0}
    \begin{tabular}{@{}lcccccc@{}}
    \toprule
    Method   &  NDCG@10 & \dm & Speedup & $k_s^{\mu}$ & $k_s^{\sigma}$\\
    \midrule
    Full              & 0.5249 & 0\%      & 1     & - & - \\
    EE$_{ideal}$      & 0.5249 & 0\%      & 3.06\x & 25.04 & 19.97 \\
    %EE$_{oracle}$ ($k=10$) & 0.5696 & +8.51\% & 7.97\x & 6.85 & 2.84 \\ da metter in tabella riassuntiva finale?
    \midrule
    ERT ($k_s = 15$)           & 0.5169 & -1.53\%  & 5.74\x & - & - \\
    ERT ($k_s = 20$)            & 0.5204 & -0.85\%  & 4.71\x & - & - \\
    EPT ($k_s = 15$, $p = 0.2$) & 0.5229 & -0.37\%  & 3.53\x & 28.56 & 12.06  \\
    EPT ($k_s = 15$, $p = 0.5$) & 0.5241 & -0.15\%  & 1.95\x & 57.49 & 31.35 \\
    \bottomrule
    \end{tabular}
\end{table}

\noindent In these experiments, EE is applied in all the cases %Techniques ERT, EPT, as well as EE$_{ideal}$, work by applying EE 
at sentinel $s=50$ %, so very early with respect to the full ensemble of $1{,}047$ trees.  
and speedup is estimated by considering as scoring cost the number of trees traversed for each document in the test set.  Note that the EE heuristic that achieves the best
%quality in the final ranking  
effectiveness -- i.e., a limited  decrease in NDCG@10 with respect to \textit{Full} ($\Delta = -0.15\%$) -- is EPT ($k_s = 15$, $p = 0.5$). Unfortunately, in this setting, also the resulting speedup is limited and lower than 2$\times$.
%but the drawback is that the speedup becomes smaller than 2x.
On the other hand, EE$_{ideal}$ obtains by construction the same NDCG@10 as \textit{Full}, but with a speedup of 3.06$\times$.
We also report in the table mean and standard deviation for the cut $k_s$ of each query ($k_s^\mu$ and $k_s^\sigma$). Obviously, speedup is inversely proportional to $k_s^\mu$, as a smaller $k_s^\mu$ implies more documents that early exit the ensemble. Note the $k_s^\sigma$ increases when we increment the EPT  score proximity $p$, because the number of documents that continue in the ensemble changes greatly for each query. This is true also for the ideal case (EE$_{ideal}$), but much less than EPT with proximity score $p=0.5$.

We can conclude saying that there is room to improve over the heuristic techniques proposed by Cambazoglu \emph{et al.} The goal is to find a trade-off between ERT ($k_s=15$), with speedups close to 6$\times$ and a very reduced ranking quality ($\Delta = -1.53\%$), and EPT ($k_s=15, p=0.5$), which obtains a very small reduction in the final ranking quality ($\Delta = -0.15\%$), but a speedup lower than 2$\times$.

%\fquestion{Non so se sia facile fare un EE$_{ideal}$ con un $\Delta \le -0.1\%$, dove scegliamo il taglio per ogni query che riduce l'NDCG della query di una max di -0.1\%.}

%\fquestion{E' possibile plottare la distribuzione dei vari tagli? cosi' potremmo forse far vedere che ci sono molti casi in cui barla piu' documenti del necessario.CLA}

%\ftodo{$EE_{ideal}$ è il migliore con il metodo di Barla, ovvero prendo i top ... alla sentinella fatto in modo da mantenere i tutti i top 10. $EE_{oracle}$ invece mantiene tutti e soli i veri top 10, non so se lasciarlo qui o in una tabella con i risultati del classificatore nella sezione dopo. Francesco}

\medskip
\noindent\textbf{\learb { Learned EArly exit Ranking.}} 
From the previous analysis, we conclude that there is space for investigating a solution that tries to improve speedup by keeping $\Delta_{NDCG@10}$  as small as possible.
%We have two possible directions, either to train a regressor able to detect the optimal cut $k_q$ for each query, thus tending to replicate the performance of $EE_{ideal}$, or to train a \textit{binary classifier} that decides for each document whether to stop its scoring or not.
%The second solution is more expensive at inference time, since we have to apply it to all the documents of a query $q$, but we experimentally verified that it works much better. \cla{\bf toglierei la prima opzione, a me suona male come regressore... al massimo lo chiamerei cmq classificatore}

We propose \lear, an ML-based solution where a binary classifier at the sentinel filters a possibly small subset of documents that are then evaluated by the remaining trees of the ensemble.
The goal is thus to discard the largest possible number of documents to boost the speedup of the ensemble computation and to select the best documents so as to provide a ranking quality as close as possible to the one achieved with the full evaluation of all candidates. These contrasting objectives pose the following challenges:
\textit{(i)} how to build the training set for the classifier;
\textit{(ii)} how to cope with imbalance of selected versus discarded documents; 
\textit{(iii)} which efficient and accurate classification algorithm to use; and, finally, 
\textit{(iv)} how to manage the trade-off between efficiency and effectiveness.
\\[.1cm]
\textit{Building the training set.} We distinguish between {\em Exit} documents and {\em Continue} documents for training the classifier. The classification label is assigned on the basis of %. To do so we exploit both information about 
the final ranking position generated by the ensemble and the relevance label associated with the document.
%, which usually consists in an integer label in the range $[1,4]$ for relevant documents and in the value $0$ for irrelevant ones. 
The set {\em Continue} includes those documents that are relevant and included in the top-$k$ results by the full ensemble.
%\textbf{QUI DOBBIAMO CAPIRE SE SCRIVIAMO CHE K=15, MOTIVANDO} 
The complementary documents define the {\em Exit} set.
At the sentinel $s$ we would like to select all the {\em Continue} documents so as to mimic the behaviour of the full ensemble.
Conversely, the {\em Exit} documents do not contribute to the top-$k$ results and their scoring should be stopped.
%it is possible to stop their traversal along the forest. 
Note that since the {\em Continue} set includes only relevant documents, and all {\em Continue} documents are, by design, ranked higher than {\em Exit} documents, a perfect classifier might drop some irrelevant documents and possibly lead to a better ranking than the full ensemble. 

To train our classifier with {\em Continue} and {\em Exit} examples
%determine a training dataset for a binary classification task. We 
we use an augmented representation for the documents including information that becomes available at the sentinel $s$. {\rp{}Specifically, besides} the features exploited by the ensemble, we use:
the rank of the document at the sentinel, 
the score accumulated up to that point, 
its per-query min-max normalized value, 
and the number of candidates for the corresponding query.
\\[.1cm]
\textit{Handling imbalance.} %We observed that the number of {\em Continue} documents is less then 1/10 of the full dataset, resulting in a highly imbalanced training set. 
{\em Continue} documents are in general a small fraction of all the instances, resulting in a highly imbalanced training set. Moreover, when using quality metrics such as NDCG@k, documents contribute differently to the quality of the result set depending on their relevance label. We tackle this issue by exploiting a cost-sensitive training where each instance $d$ having relevance label $r_d$ and classification label $l_d$, is associated with a different weight $w_d=2^{r_d} /f_q(l_d)$, where$f_q(l_d)$ is the frequency, among the candidates for query $q$, of the {\em Continue}/{\em Exit} classification label $l_d$. This pushes the classifier to prioritize loss reduction on documents with large relevance labels proportionally to their contribution to the NDCG metric, and on the infrequent {\em Continue} documents. 
Note that this weighting scheme is query-based and  allows the classifier to adapt to the different queries in the dataset.
\\[.1cm]
\textit{Classifier efficiency.} Several options are available for building a binary classifier, e.g., logistic regression, SVM, etc. Note that the classification task performed for each document is a potential overhead that we are introducing.
After experimental evaluation, not included in this work due to space constraints, we chose a small forest of $10$ trees trained by minimizing the logistic loss. We found that such a small forest provides the best results with a limited additional cost.
\\[.1cm]
\textit{Efficiency vs. effectiveness trade-off.} Accuracy is not the metric we are targeting for the classifier. Our goal is in fact to maximise the recall over {\em Continue} documents, without hindering precision.
%by minimizing\rp{, at the same time,}  the total number of documents guessed as .
To this end, we fine tune a filtering threshold on the probability of belonging to class {\em Continue} predicted by the classifier. By varying this threshold, we can find the sweet spot between precision and recall.
Finally, the position of the sentinel $s$ impacts on the accuracy of the classifier and on the efficiency of the \lear framework. Early sentinels generate less reliable partial document scores/ranks (due to the limited number of trees) potentially harming the classifier accuracy. On the other hand, 
%they may produce large speedups thanks to the anticipated exit of several documents.
they may produce large speedups thanks to the amount of tree traversals avoided.
In the experimental section we investigate different sentinel points.

\section{Experiments}
\label{sec:experiments}
\noindent \textbf{Datasets}.
The datasets used for experiments are \msna\footnote{\url{http://research.microsoft.com/en-us/projects/mslr/}} (Fold 1) and  \istella\footnote{\url{http://blog.istella.it/istella-learning-to-rank-dataset/}}. The \msna one consists of $31$,$351$ queries and $136$ features extracted from $3$,$771$,$125$ query-document pairs, while the \istella dataset is composed of $33$,$018$ queries and $220$ features extracted from $10$,$454$,$629$ query-document pairs. They thus differ in the average number of documents per query, ranging from the $120$ of \msna to the $317$ of \istella. The query-document pairs in both datasets are labeled with relevance judgments ranging from $0$ (irrelevant) to $4$ (perfectly relevant). \istella comes with about $96$\% of non-relevant documents and a normal distribution among the relevant ones centered on label 2, while \msna shows a power law distribution with 51\% of non-relevant documents.
%To this regard, another big difference between the two datasets is that \istella comes out with about $96$\% of non-relevant documents and a normal distribution centered on label 2 documents among the relative ones, while \msna shows a power law distribution with 51\% of non-relevant documents.
%
Both the datasets are split in four partitions with sizes 60\%-20\%-5\%-15\%: the first partition is used to train 
the \lmart ranking model; the second for hyper-parameter tuning of \lmart; the binary classifier is trained on the second partition and fine-tuned on the third; finally, the fourth partition is used as test set to evaluate the efficiency and effectiveness of the ensemble and the EE strategies considered.

%i) $\texttt{train}_\texttt{ml}$ for training the \lmart ranking model, ii) $\texttt{vali}_\texttt{ml}$ = $\texttt{train}_\texttt{lear}$ for \lmart hyper-parameter tuning and and for training the \lear classifier, iii) $\texttt{vali}_\texttt{lear}$ for \lear hyper-parameter tuning, and iv) $\texttt{test}_\texttt{ee}$ for testing the efficiency/effectiveness of \lear compared to state-of-the-art early stopping solutions. The latter two partitions are split accordingly to a 25\%/75\%  scheme from the original test set of the corresponding datasets.

% The \msna dataset is split in train, validation and test sets according to a 60\%-20\%-20\% scheme, while the \istella dataset is distributed with only train/test splitting according to a 70\%-30\% scheme. For the purpose of this article we randomly selected 20\% of the dataset queries from the train and build a validation set, resulting thus in a 50\%-20\%-30\% scheme  of train, validation and test sets splitting.

%\ftodo{discutere il competitor: BARLA e la versione usata. come e' determinato il threshold. ecc.}
%\ftodo{loro propongono 4 alternative --> migliore è EPT (proximity based) ovvero tengo i primi k alla sentinella e quelli vicini al k-esimo--> poco impatto, threshold deciso a priori, nessuna anlisi qui, il nostro confronto lo facciamo con thresholds da 0.1 a 0.9 a step di 0.1, gli stessi del classificatore}

%\vspace{1mm}
\noindent \textbf{Ranking models}.
The reference ranking models were trained with \lmart. 
%, a state-of-the-art gradient boosting algorithm that learns an additive ensembles of regression trees exploiting NDCG in the loss function. 
Indeed, we used the LightGBM implementation\footnote{\url{https://github.com/microsoft/LightGBM}}, and fine-tuning hyper-parameters by maximizing NDCG@10 and using a bayesian approach as provided by HyperOpt\footnote{\url{https://github.com/hyperopt/hyperopt}}. 
%The hyper-parameters tuned are: learning rate, number of leaves, max depth, minimum number of instances in leaves, minimum sum of hessian, minimum gain to split
The number of trees was limited at $1$,$500$ and tuned with 100 iterations for early stopping. The resulting ensembles have $1$,$047$ and $1$,$469$ trees for \msna and \istella, respectively. Each tree has up to $64$ leaves. %The MSN-1 ensemble is the one  already discussed in Section \ref{sec:learning}.

%\vspace{1mm}
\noindent \textbf{Competitor algorithm}.
We evaluate \lear against the EPT heuristic strategy.
For EPT we used $k_s=15$ and proximity thresholds $p$ ranging from $0.3$ (more aggressive EE) to $0.8$ (more conservative EE), with a step size of $0.1$.

%\vspace{1mm}
\noindent \textbf{\textsc{LEAR} binary classifier}.
As previously discussed, we use a forest of 10 trees to detect {\em Continue}/{\em Exit} documents. We used top-$15$ results ranked by the full ensemble to determine the document in the two classes.
The model was trained by optimizing the logistic loss and using the same implementation framework of \lmart (LightGBM + HyperOpt). 
%We trained \lear on tree different sentinels, i.e., $s=\{50,100,200\}$.
%We employ a tree-based classifier to perform our early exit classification in \lear. In detail, the classifier is an ensemble of trees that optimizes binary logloss on the training set. The learning setting is similar to that of \lmart (LightGBM + HyperOpt). Since the classifier has to introduce a limited overhead in the scoring process with EE,  we severely limit the number of trees up to $10$. 
We also performed  feature importance analysis on query-document features. The resulting models employs $54$ features for \msna and $118$ for \istella. 
%To further reduce the cost of the binary classifier, after feature importance analysis, we reduced the number of considered feature to $54$ for \msna and $118$ for \istella.
%\cla{{\bf Tutto questo va messo dopo.} {\bf DICIAMO CHE SE IL TITOLO FOSSE "ML-Framework for EE" invece di "\lear Binary Classifier", ci sta anche la costruzione dei training set. CLA: il training set lo abbiao già spiegato prima, il variare di s mi pare specifica degli esperimenti successivi}
%We trained \lear by constructing three distinct training datasets, one for each sentinel $s=\{50,100,200\}$, by using the learnt \lmart model applied to second partition of \msna and \istella. Therefore, the learnt classifier changes depending on the sentinel position.
Finally, in all tests for \lear we used different choices of confidence classifier thresholds, ranging from $0.1$ (more conservative EE) to $0.7$ (more aggressive EE), with steps of $0.1$. Note that the effect of these thresholds is similar to the one modeled by parameter $p$ in EPT.

\noindent \textbf{Assessing the efficiency}.
We assess the efficiency of \lear and \ept using \QS (QS) \cite{quickscorer15,Dato2016}, the state-of-the-art algorithm for scoring ensembles of regression trees. We extend QS by introducing the computation of early exit strategies (both \lear and \ept) at a given sentinel during the scoring process. All the results reported hereinafter %are computed by 
consider the total latency of the process, i.e., 
the time needed to score the documents with the \lmart model plus the time needed to compute the early exit strategy. Our extended QS is implemented in C++. %\sal{The algorithm traverses the ensemble in blocks of trees to optimize memory hierarchies usage, and block boundaries is a suitable point where a sentinel can be placed to activate our early-exit classifier.}
%The code is compiled with GCC 5.4.0 with the \texttt{-O3} optimization flag. 
%The experiments have been performed in single-thread execution. Scoring times are measured on a Intel Xeon E5-2650 v3 CPU (@3.0 GHz) with 128 GiB of RAM, a L1-cache of 32 KiB, a L2-cache of 256 KiB, and a L3-cache of 25 MiB.

% !TEX root = paper.tex
% !TeX spellcheck = en_US

\subsection{Performance of \learb classifier}

Table~\ref{tab:class-metrics} reports the precision and recall performance of the \lear classifier. 
%on the two  datasets used. 
\lear exhibits a very large recall for the {\em Continue} class of 97\% and 99\% on the \msna and \istella datasets, respectively. Therefore, %is high figures show that 
the classifier is able to identify nearly all documents that should continue the forest traversal because they are likely to be included into the top-$k$ results. Having such a large recall on the {\em Continue} class is necessary for a high-quality final ranking. 
The second objective of the classifier is to minimize the number of false positives, i.e., the number of {\em Exit} documents incorrectly classified as {\em Continue}, as they do not contribute to the final top-$k$ results and only increase the overall evaluation cost. In this regard,
the recall on the {\em Exit} class is 82\% for \msna and 91\% for \istella, which are pretty good results given the large amount of documents which we expect to prune.  
%This signifies that more than 80\% of the irrelevant documents are discarded by \lear at the sentinel. We recall that in both datasets the irrelevant documents are more then 90\% of the total. 
These results were achieved with a sentinel $s=50$, by thresholding the classifier's predicted probability at 50\%. In the following subsection, we investigate the impact on performance of these two tuning parameters.

\begin{figure}[]
    \centering
    \includegraphics[width=0.85\columnwidth]{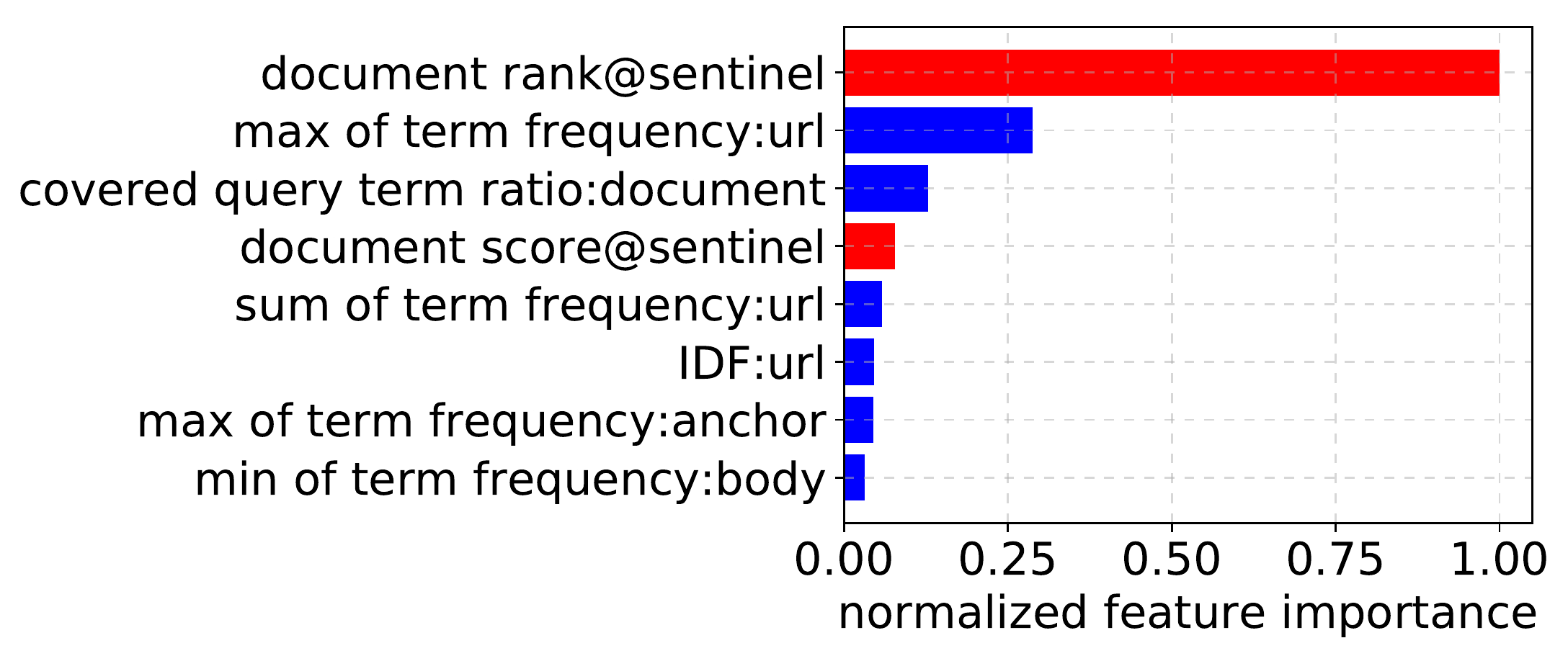}\vspace{-3mm}
    \caption{Feature importance analysis.\label{fig:importance}}
    \vspace{-4mm}
\end{figure}

Figure\ref{fig:importance} reports the feature importance analysis of the binary classifier including the rank- and score-based features made available at the sentinel. Notably, the document rank and the document score at the sentinel (red bars) are the first and fourth most important features, thus largely contributing to the classification accuracy.

\subsection{Efficiency/effectiveness trade-off}
\begin{figure}[b!]
    \centering
    \begin{subfigure}[b]{0.495\columnwidth}
        \centering
        \includegraphics[width=\columnwidth]{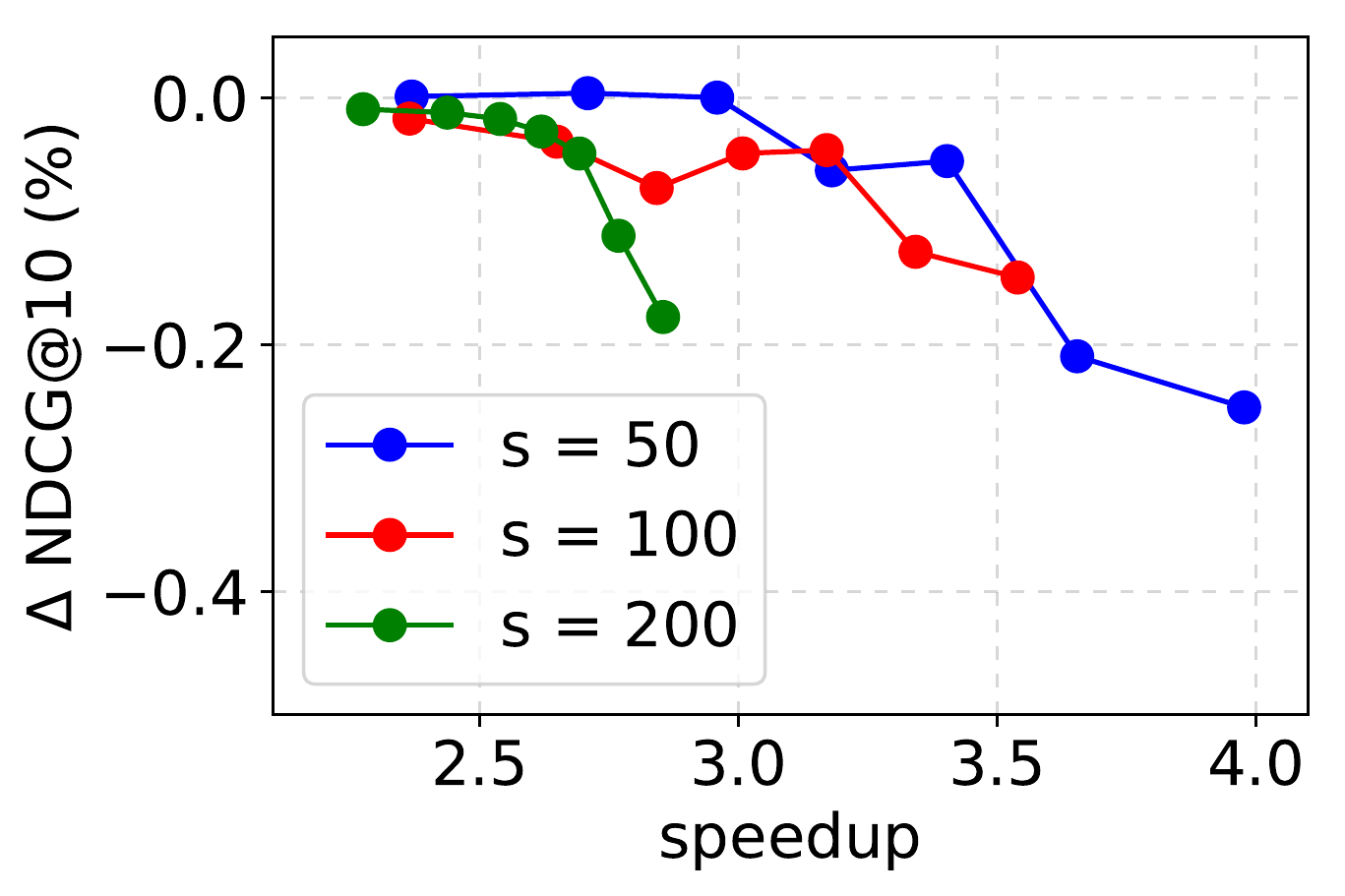}
        \vspace{-3mm}
        \caption{\learb\label{fig:001}}
    \end{subfigure}
    \begin{subfigure}[b]{0.495\columnwidth}
        \centering
        \includegraphics[width=\columnwidth]{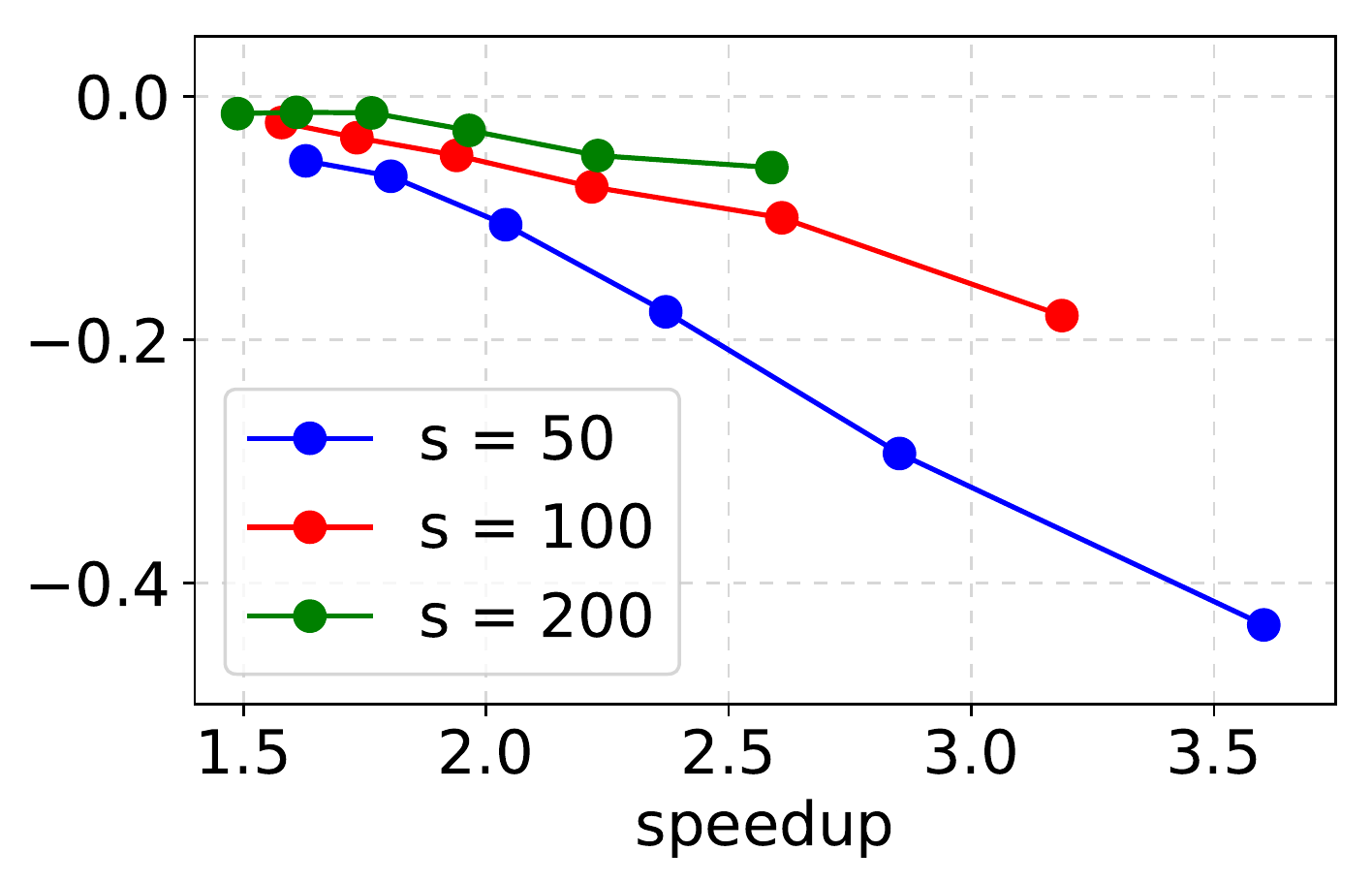}
        \vspace{-3mm}
        \caption{\eptb\label{fig:002}}
    \end{subfigure}
    \vspace{-7mm}
    \caption{Efficiency-effectiveness trade-offs of \learb and EPT by varying the sentinel on the \msnab dataset.\label{fig:tradeoff-msn}}
\end{figure}

We compare the performance of \lear against \ept by evaluating the efficiency/effectiveness trade-offs for the \msna dataset.
The two plots in Figure \ref{fig:tradeoff-msn} show the impact of the two strategies on: \textit{i)} the final ranking quality, evaluated in terms of reduction of NDCG@10 compared to the reference \lmart model  (y-axis), and \textit{ii)} the speed-up derived from the reduced number of documents to be scored through the full ensemble (x-axis).
%The two plots show the effectiveness of the two methods in terms of reduction of NDCG@10 caused by the introduction of the two techniques (y-axis) by varying their efficiency \sal{(x-axis).} % Specifically, for both EPT and \lear we can increase/decrease the aggressiveness of EE by tuning their threshold parameters, thus increasing/decreasing their efficiency and thus the overall speedup.}  
%ity  i.e., the overall speedup achieved by the query processor (x-axis) when employing them. 
Specifically, each plot reports three curves, each corresponding to the performance of the method when applied at a given sentinel, i.e., after $50$, $100$, and $200$ trees of the \lmart model. For each sentinel, the curve reports the different efficiency/effectiveness trade-offs obtained by varying the confidence/proximity threshold for \lear/\ept, respectively. The two plots allow us to easily identify the sentinel with the best efficiency/effectiveness trade-off, i.e., the line dominating the others in terms of final ranking quality and achieved speedup.
Figure \ref{fig:tradeoff-msn} (a) shows that the best choice for \lear is to put the sentinel after the $50$-th tree. In this setting, \lear achieves no effectiveness degradation with a speedup of up to 3$\times$ for small values of the confidence threshold (up to $0.3$).

\begin{table}[t!]
%    \small
    \centering
    %\ra{1.0}
    \caption{\lear Precision and Recall figures.\label{tab:class-metrics}}
    \begin{tabular}{@{}lccccc@{}}
    \toprule
    \multirow{2}{*}{Class} & \multicolumn{2}{c}{\msna} & \multicolumn{2}{c}{\istella} \\
    \cmidrule(lr){2-3}
    \cmidrule(lr){4-5}
    & Precision & Recall & Precision & Recall \\
    \midrule
    Exit & 1.00 & 0.82 & 1.00 & 0.91  \\ % 90 95
    Continue & 0.33 & 0.97 & 0.25 & 0.99  \\ % 49 40
    \bottomrule
    \end{tabular}
    \vspace{-3mm}
\end{table}

\begin{figure}[b!]
    \centering
    \begin{subfigure}[b]{0.49\columnwidth}
        \centering
        \includegraphics[width=\columnwidth]{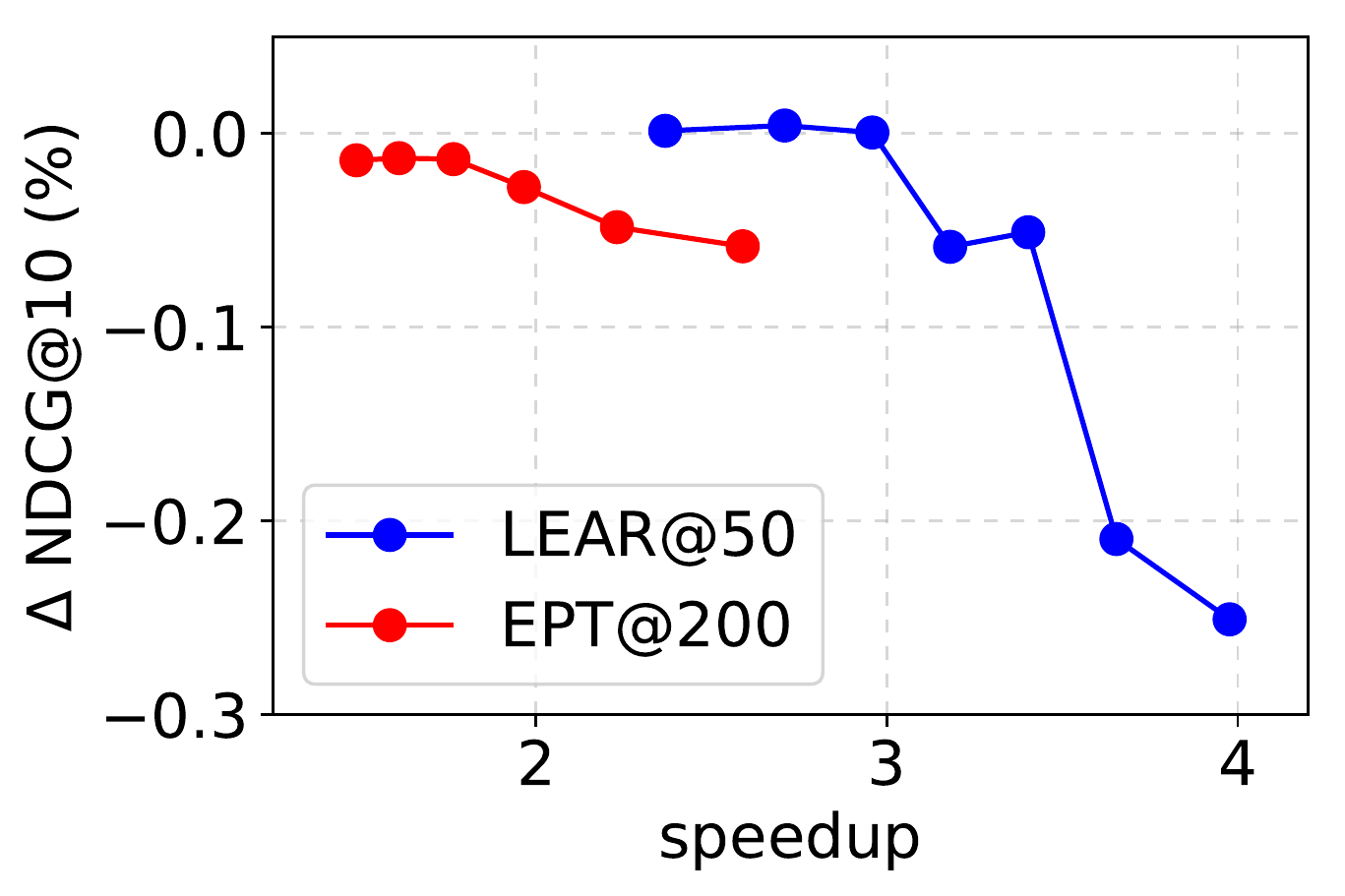}
        \vspace{-3mm}
        \caption{\msnab\label{fig:001b}}
    \end{subfigure}
    \begin{subfigure}[b]{0.49\columnwidth}
        \centering
        \includegraphics[width=\columnwidth]{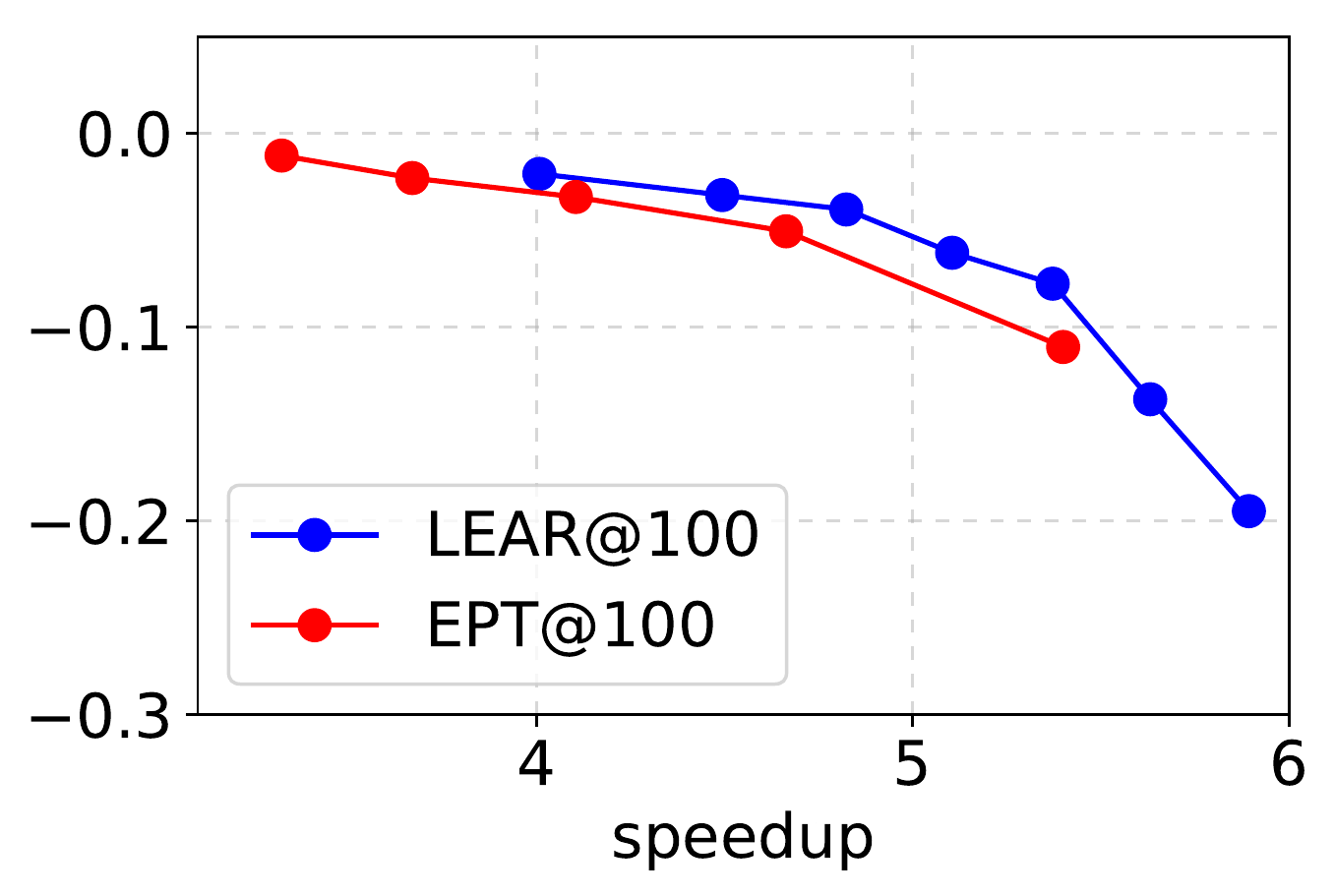}
        \vspace{-3mm}
        \caption{\istellab\label{fig:002b}}
    \end{subfigure}
    \vspace{-3mm}
    \caption{Efficiency-effectiveness trade-offs of the best \learb against the best EPT on the \msnab and \istellab datasets.\label{fig:tradeoff-compared}}
\end{figure}

Note the this result exactly corresponds to that of EE$_{ideal}$, reported in Table \ref{tab:motivation} as the best possible result to which EPT can tend. 
By increasing the \lear confidence threshold, the EE strategy becomes more aggressive. This translates to higher speedups at the cost of higher degradation of the ranking quality. 
Similarly, Figure \ref{fig:tradeoff-msn} (b) reports the same analysis for \ept by varying the proximity threshold. Here, the best trade-offs are achieved at $200$ trees. In this setting, results show no quality degradation with speedups of up to 1.75$\times$ by using proximity thresholds higher than $0.6$. From that point, in correspondence of lower proximity values, \ept starts degrading ranking quality with a maximum loss observed of $-0.05\%$ in correspondence to a speedup of about 2.6$\times$. On the \msna dataset, \lear and \ept thus show their best results by using sentinels placed at different points of the ensemble. We also tested the two methods on the \istella dataset. We do not report the plots due to space constraints. Results on \istella show that the best performance of the two methods are achieved at $100$ trees. 
We conclude the analysis by reporting a direct comparison of the best sentinel placements of the two methods in Figure~\ref{fig:tradeoff-compared}. Figure \ref{fig:tradeoff-compared} (a) provide such a comparison on the \msna dataset. It clearly show the superiority of \lear with respect to \ept. Indeed, the former method outperforms the latter by margin, providing much higher speed-ups when keeping fixed the effectiveness degradation or, conversely, providing higher effectiveness at the same speed-up ratio. Figure \ref{fig:tradeoff-compared} (b) shows the same comparison on \istella, where \lear still outperforms \ept in terms of trade-offs despite a reduced margin. To conclude, we experimentally show that our ML-based solution for early exit additive ranking ensembles is able to achieve better efficiency-effectiveness trade-offs than previous state-of-the-art heuristics.

% !TEX root = paper.tex
% !TeX spellcheck = en_US

\section{Conclusion}% and Future Work}
\label{sec:conclusions}
We have discussed \lear, an effective ML-based technique to speedup document ranking employing additive ensembles of regression trees. \lear forces documents to early exit the ensemble if they are unlikely to be ranked among the final top-$k$ results.
Experiments on two public datasets showed that \lear achieves speedups larger than 5$\times$ with a negligible loss of NDCG@10 ($<0.05$\%). Results also showed that it remarkably outperforms the state-of-the-art document-level EE heuristics.
As future work, we intend to investigate the integration in the \lear framework of query-level early exit strategies.

%\noindent\textbf{Acknowledgements.} Work partially supported by projects BIGDATAGRAPES (EU Horizon 2020, grant No. 780751)  and  OK-INSAID (MIUR, grant No. ARS01\_00917).

\newpage
\bibliographystyle{ACM-Reference-Format}
\bibliography{biblio}

\end{document}